\documentclass[12pt]{article}

\newcommand{\m}{\mathrm}
\newcommand{\be}{\begin{equation}}
\newcommand{\ee}{\end{equation}}
\newcommand{\ba}{\begin{eqnarray}}
\newcommand{\ea}{\end{eqnarray}}

\usepackage{graphicx}
\usepackage{amssymb}
\usepackage{amsmath}
\usepackage[T1]{fontenc} 
\usepackage[ansinew]{inputenc} 
\usepackage[nosort]{cite}
\newcommand{\inbar}{\vrule height1.57ex width.4pt depth0pt}
\newcommand{\SW}{\relax{\hbox{$\ \inbar\kern-.285em{\rm S}$}}}

\begin{document}
\thispagestyle{empty}
\begin{center}

\null \vskip-1truecm \vskip2truecm

{\Large{\bf \textsf{Evidence that the Rate of Evolution of a Black Hole Interior Has a Holographic Dual}}}

{\large{\bf \textsf{}}}

{\large{\bf \textsf{}}}

\vskip1truecm

{\large \textsf{Brett McInnes}}

\vskip1truecm

\textsf{\\  National
  University of Singapore}

\textsf{email: matmcinn@nus.edu.sg}\\

\end{center}
\vskip1truecm \centerline{\textsf{ABSTRACT}} \baselineskip=15pt
\medskip

A ``large'' AdS black hole can attain equilibrium with its own Hawking radiation, and in that condition it is thought to be dual to a strongly coupled field theory, also at equilibrium. But the interior of the black hole is by no means static: the geometry of spatial sections lying inside the event horizon evolves at some rate. This prompts the obvious question: can this rate possibly have a holographic dual? We present circumstantial evidence that such a dual does exist. We do this by making concrete proposals for two objects: first, a rough measure (already suggested in the literature) of the rate at which the unitary evolution of the state vector describing equilibrated strongly coupled matter evolves; and, second, a rough measure of the rate at which the interior (just under the horizon) evolves (in the case of the AdS$_5$-Kerr black hole). We then study how these two very different objects change as two physical parameters describing the exterior of the bulk black hole (the specific angular momentum and the temperature) are varied. We find that they change in remarkably similar ways, as holographic duals should.

\newpage

\addtocounter{section}{1}

\addtocounter{section}{1}
\section* {\large{\textsf{1. ``Time-Dependent Thermal Equilibrium'' ?}}}
The exteriors of asymptotically Anti-de Sitter five-dimensional black holes have long occupied centre stage in the study of the gauge-gravity duality \cite{kn:nat}. Recently, however, attention has naturally turned to the \emph{interiors} of these objects.

The primary novelty here is this: whether interpreted in terms of evolving volumes \cite{kn:oyc,kn:interior,kn:anything3}), or classically pinching (yet possibly quantum-traversable) wormholes \cite{kn:aronwall,kn:juan,kn:bilotta}, these interiors are highly \emph{dynamic} spacetimes; and this is true even after the exterior has settled down to a state of thermal equilibrium, which is possible when the black hole is ``large'' (as we shall always assume). The boundary theory, which describes a form of strongly-coupled matter, is likewise at thermal equilibrium, and therefore apparently static. This prompts the obvious question: how can a static boundary theory give a holographic account of a time-dependent black hole interior? Resolving this apparent paradox \emph{presents a serious challenge for bulk-boundary duality.}

This has been a problem for AdS/CFT duality from the outset. The obvious way to deal with it would be by means of an explicit extension of the duality to the black hole interior. There have of course been many attempts to do so \cite{kn:oog}, but the question remains open: see \cite{kn:cheth} for a clear recent review.

It may be that the AdS/CFT duality only relates the black hole \emph{exterior} to the boundary physics. That would be disappointing indeed. Furthermore, for those who believe that the event horizon is a mere locus, not some kind of barrier (``no drama at the event horizon'') it is also highly implausible: why should the remit of the boundary physics be suddenly cut off at spacetime points where nothing happens when they are traversed? Here we will assume that there is no drama at the event horizon, so that, in fact, the geometry just under the event horizon is a simple continuous extension of the geometry just above it, which in turn is controlled by physics that does have a holographic dual. For us, therefore, the apparent paradox is particularly acute.

Now the quantum description of thermodynamic equilibrium differs from its classical counterpart in one way: the unitary evolution of the state vector does not ``notice'' that classical equilibrium has been attained ---$\,$ it continues unabated \cite{kn:scott}. Normally we attach no great physical significance to this fact, but, in view of the above, \emph{perhaps we need to revise this assumption in the case of strongly coupled matter}. That is, we argue that the continuing ``post-equilibrium'' unitary evolution does have some physical meaning in this particular case, and that (in some sense we will have to make more precise) \emph{the rate at which this occurs is holographically dual to the rate at which the dual black hole interior evolves}. Finding evidence for this assertion is the goal of the present work.

As is well known, Susskind and others \cite{kn:suss1,kn:suss2,kn:suss3,kn:suss4} have proposed a specific implementation of these ideas, in terms of the \emph{quantum circuit complexity} of strongly coupled matter, as represented by the boundary field theory. See \cite{kn:poli} for a recent review.

While there is much to be said in favour of this proposal, it remains controversial. Complexity is defined quasi-teleologically in terms of minimal numbers of gates which must be executed to attain a desired state, starting at some specified initial state. There is no definite prescription for how the gates are chosen, and, more generally, it is far from easy to see how such a quantity can measured or indeed be ``physical'': see for example \cite{kn:umesh,kn:sarosi} for discussions.

\emph{Here we will completely avoid this controversy}, and focus instead purely on the independently very interesting claim that the post-equilibrium unitary evolution of strongly coupled matter ---$\,$ which, let us remember, is itself far from fully understood ---$\,$ occurs at a certain rate $\mathfrak{R}$ which is in some sense \emph{also} the rate of change of the spatial geometry (just) inside the dual bulk black hole.

The key point here is that one can argue that $\mathfrak{R}$ can be (roughly) expressed in physical terms on the boundary even if one does not fully understand the physical meaning of the underlying process. If the process in question is carried out, as seems natural, by thermal fluctuations, then a simple argument (advanced in \cite{kn:jacob}) shows that its rate should, in the simplest cases\footnote{In practice, ``in the simplest cases'' means that the bulk black hole is neither extremal nor close to extremality.}, be given (again, roughly) by $\mathfrak{s}\,T/\hbar$, where $\mathfrak{s}$ is the specific entropy\footnote{Throughout this work, Gothic lettering is used to denote thermodynamically \emph{intensive} quantities.}, $T$ is the temperature, and $\hbar$ is the reduced Planck constant: see \cite{kn:jacob,kn:tallarita} for more detail.

If $\mathfrak{R}$ is given by this explicitly ``physical'' expression, then we can investigate how it varies as we change the parameters describing strongly coupled matter. In fact we can do this in two ways: first, by examining phenomenological models of strongly coupled matter, and second, by using the conventional holographic duality of the exterior black hole geometry with the physics of strongly coupled matter on the boundary. We find that, to the (admittedly very limited) extent that quantitative phenomenological models relevant to our specific concerns have been developed to date, the two methods agree.

For example, a phenomenological model of the Quark-Gluon Plasma\footnote{This is the form of strongly coupled matter for which we have experimental data and phenomenological models: it is produced in heavy-ion collisions \cite{kn:STAR,kn:franc,kn:prefut,kn:hot,kn:sign}. The long-standing hope has been that at least some properties of strongly coupled matter are universal, so that the study of the QGP might teach us some (qualitative) lessons regarding the (very different) kind of strongly coupled matter that is directly dual to an AdS$_5$ black hole.} (QGP) indicates, as we will see, that $\mathfrak{s}\,T/\hbar$ should increase with $T$ when all other parameters are fixed, \emph{but} also, more unexpectedly, that it is bounded above as $T$ is taken to be arbitrarily large. We find that the holographic dual of $\mathfrak{s}T/\hbar$, where now $\mathfrak{s}$ is the specific entropy of the black hole, and $T$ is its Hawking temperature, behaves in much the same way. Still more striking, we find that similar comments (probably) apply when the temperature is fixed and the other key parameter describing the (``vortical'') QGP, namely the specific angular momentum $\mathfrak{j}$ \cite{kn:jiang}, is varied. Thus we have a fairly good understanding of the parameter-dependence of $\mathfrak{R}$, from two very different points of view.

The key point now is that, assuming ``no drama'' as above, the exterior geometry of the black hole, just above the event horizon, gives us access to the region inside the black hole, just below the event horizon. Again, we propose a concrete expression for a measure of the rate of evolution of the internal spatial sections just below the event horizon, and we can study how it varies as $T$ and $\mathfrak{j}$ are changed. Using an AdS$_5$-Kerr geometry, we find good agreement with both (phenomenological and holographic) of our studies of $\mathfrak{R}$. This provides good circumstantial evidence that $\mathfrak{R}$ is in fact the holographic dual of the rate at which the black hole interior (again, just below the event horizon) changes.

It should be stressed that our procedure here differs very substantially from the one followed in the recent literature on holographic duality for AdS black hole interiors (see for example \cite{kn:anything3}). Here we are working out the duality in two steps: boundary to bulk exterior to bulk interior. Notice that the holographic part of our argument involves quantities, $T$, $\mathfrak{s}$, and $\mathfrak{j}$, all of which can be determined by observations (of frame-dragging, and so on) taken \emph{outside} the black hole.
For us, therefore, it is essential to keep the black hole interior strictly segregated from the exterior, and so, when we come to foliate the interior, we do so with slices which never penetrate the event horizon: they remain strictly inside. Thus, in particular, we have nothing to say regarding the ``complexity = volume'' and allied conjectures.

We should note that some of the ideas discussed here were also considered in \cite{kn:110}. There, however, we used a different identification of the physical mass of an AdS$_5$-Kerr black hole, which, in view of the recent important work reported in \cite{kn:gaogao}, we now consider to require revision. Adopting the identification proposed in \cite{kn:gaogao} radically changes (and complicates) the entire analysis and also many of the results, even in a qualitative sense. Some of the results of \cite{kn:110} do survive, however, and we discuss this in the Conclusion.

\addtocounter{section}{1}
\section* {\large{\textsf{2. $\mathfrak{R}$ in Terms of Specific Entropy and Temperature}}}

If ``something is happening'' in strongly coupled matter at thermal equilibrium, presumably this is being performed by thermal fluctuations of the \emph{active} degrees of freedom in that matter. In theory (see \cite{kn:jacob} for this argument) one would count these local degrees of freedom using the (total) entropy of the system. However, in practice, the total entropy is never used in studies of actual plasmas such as the QGP, because the full extent of the plasma formed in a collision varies from one collision to another, and indeed is not sharply defined. For this reason we always use \emph{specific} quantities (specific entropy and specific angular momentum); and in fact there are also strong reasons, when we come to the holographic interpretation (see below), to prefer such quantities. Doing this just means that the rate we are about to compute is the rate of change of some thermodynamically intensive quantity.

Thus we count the active degrees of freedom using the specific entropy $\mathfrak{s}$, the ratio of the entropy density to the mass density. (To be precise, the count is given by $\mathfrak{s}/k_{\textsf{B}}$, where $k_{\textsf{B}}$ is the Boltzmann constant.)

The relevant time scale here\footnote{This time is measured by an observer who rotates with the plasma, since that is the time associated with equilibration. It is not the same as the time measured by the stationary observer at the pole, but nor is the temperature the same. It is shown in  \cite{kn:jacob} that these two effects \emph{cancel} when we compute the rate, so we can interpret time and the temperature as those measured by an external observer. In particular, $T$ is still the Hawking temperature.} is the characteristic time scale of thermal fluctuations. That is, it is set by the reciprocal of the temperature $T$: to be precise, by $\hbar /\left(k_{\textsf{B}}T\right)$. We therefore suppose that whatever is happening in strongly coupled matter at equilibrium happens at a rate proportional to the ratio of $\mathfrak{s}/k_{\textsf{B}}$ to $\hbar /\left(k_{\textsf{B}}T\right)$; and so we have
\begin{equation}\label{ALPHA}
\mathfrak{R} \;=\; {\mathfrak{s}\,T\over \hbar}.
\end{equation}
The expression on the right has the appropriate units, those of 1/(mass $\times$ time).

Clearly this argument is drastically over-simplified: we do not really believe that the rate of change of the state of so complex a system as the QGP can be captured by such a simple expression. Instead we regard $\mathfrak{s}\,T/\hbar$ as a coarse average over the many parameters no doubt required to give a full picture of the time-dependence of the state. We interpret the equality in (\ref{ALPHA}) as agreement up to a dimensionless factor of order unity \cite{kn:tallarita}. (Later we will argue that $\mathfrak{s}\,T/\hbar$ is a good measure of the rate, but only at relatively early times.)

Notice that $\mathfrak{R}$ is an intensive quantity\footnote{If one were to accept the identification of $\mathfrak{R}$ in terms of complexity, then, to be exact, it would be the rate of change of the \emph{specific} complexity, or ``the complexity allowing for the resources available to execute gates''. This would seem to be an interesting quantity to consider in any case.}, whence the notation. Notice too that (as we will see in detail later in both phenomenological and holographic models) $\mathfrak{s}$ depends on both $\mathfrak{j}$ \emph{and} $T$, so the $T$-dependence of $\mathfrak{R}$ is not obvious.

Much of our subsequent work will be devoted to the study of this apparently simple quantity.

\addtocounter{section}{1}
\section* {\large{\textsf{3. About the Vortical QGP}}}
The temperature of the QGP has of course always been of great interest. A more recent realisation was that it is actually possible to observe signatures of \emph{vorticity} in the QGP (at the RHIC facility \cite{kn:STARcoll,kn:STARcoll2}). This is correlated with the centrality of the collision experimentally, and with the specific angular momentum $\mathfrak{j}$ of the resulting plasma theoretically. This has led to an explosion of theoretical work on vortical QGP plasmas: see for example \cite{kn:becca} or more recently \cite{kn:fei}. In short, $\mathfrak{j}$ is now as interesting, experimentally and theoretically, as $T$.

However, it is still the case that most of the phenomenological literature on the QGP is focused on low-centrality collisions: see for example \cite{kn:sahoo}, which is designed to reproduce various properties of the QGP produced in collisions (at up to 10$\%$ centrality, that is, collisions resulting in relatively low angular momentum densities) at the RHIC facility. Let us begin, then, by asking how $\mathfrak{s}\,T/\hbar$ varies in such a model, when $\mathfrak{j}$ is fixed at some small value, while the impact energy varies. The relevant results are shown in the table below\footnote{We follow \cite{kn:sahoo} and use particle theory units in this Section (only), so here mass is measured as multiples of MeV/$c^2$, while time is measured as multiples of $\hbar$/MeV, so (since $\mathfrak{R}$ has units of 1/(mass $\times$ time) if we wish to convert the entries in the right column of the Table to conventional units, we just have to regard them as multiples of $c^2/\hbar$.}.

\begin{center}
\begin{tabular}{|c|c|}
  \hline
$\sqrt{s_{\m{NN}}}$(GeV)&$\displaystyle \mathfrak{s}\,T$ \\
\hline
$7.7$ &  1.096 \\
$11.5$ & 1.120 \\
$14.5$ &  1.158 \\
$19.6$ &  1.165 \\
$27$  &   1.191 \\
$39$  &   1.196 \\
$62.4$  &  1.235 \\
$200$  &  1.237 \\
\hline
\end{tabular}
\end{center}

Here $\sqrt{s_{\m{NN}}}$ is the impact energy per nucleon pair. The values steadily increase, but ever more slowly: even the large step in $\sqrt{s_{\m{NN}}}$ from $62.4$ to $200$ GeV has almost no effect. Clearly $\mathfrak{s}\,T$ is bounded above, and is asymptotic to some value. The thermodynamic Euler relation indicates what that asymptotic value must be: in the ultra-relativistic limit, the pressure is nearly one third of the energy density, so in the most extreme cases the Euler relation implies $\mathfrak{s}\,T \approx 4/3$ if the baryonic chemical potential can be neglected (as it can be in LHC plasmas). There is reason to think that this value \emph{is indeed approached} in central LHC collisions \cite{kn:sign}.

Values below about $1$ do not occur, at least in central collisions, because the QGP does not form in the relatively low-energy collisions needed to produce such values in the central case. Thus, in the actual QGP produced in central or near-central collisions, $\mathfrak{s}\,T$ is a monotonically increasing function with values confined to a narrow range between about $1$ and just over $1.3.$ We will see later that a simple holographic model reproduces this range surprisingly well.

Less apparent from the data is the fact that all of these values of $\mathfrak{s}\,T$ are very large by ordinary standards (because $c^2/\hbar$ is large in conventional units, around $8.52 \times 10^{50}/(\m{kg}\cdot \m{s}).$ The value for water at 300 K is about 11 orders of magnitude smaller \cite{kn:110} (see \cite{kn:nist} for the data). If indeed $\mathfrak{s}\,T$ measures a rate of change, then, the QGP produced in the highest-energy collisions is probably the experimentally accessible form of matter which changes most rapidly in this sense. Thus indeed the QGP is probably the right medium to use to investigate our ``paradox''.

Now let us turn to the other case, in which $T$ is fixed but $\mathfrak{j}$ is not negligible. Phenomenological models of the QGP with significant vorticities are only beginning to appear. In \cite{kn:kshitish} such a model is described which predicts that angular velocities (at given, fixed temperatures) up to around 2 fm$^{- 1}$ tend to produce a mild \emph{increase} in $\mathfrak{s}\,T$ (computed by using the data in Figure 1 of \cite{kn:kshitish}). For example, if $T \approx 180$ MeV, then $\mathfrak{s}\,T$ increases from approximately $1.196$ at low angular velocities (compare with $\sqrt{s_{\m{NN}}} = 39$ GeV in the Table above) to about $1.277$. Thus we can say tentatively that fairly large specific angular momenta tend to increase $\mathfrak{R},$ at fixed temperature, though not by a large factor.

While this is a relatively small increase compared to $1.196$, note however that $1.277$ is higher than any value attained in central collisions at the RHIC, according to the model discussed in \cite{kn:sahoo}. We will argue later that large angular momenta permit values of $\mathfrak{s}\,T$ higher than \emph{any} value possible in central collisions (that is, higher than $4/3$ as discussed above). \emph{Thus angular momentum can affect $\mathfrak{s}\,T$ in ways that have characteristic experimental signatures.}

Very recently, the case of even larger vorticities (perhaps produced in the plasmas observed in the ALICE facility at the LHC \cite{kn:prefut,kn:hot}) has begun to attract attention \cite{kn:buzztuch} (see also \cite{kn:sarthak}). In these very extreme cases, the angular velocity, $\omega$, is so large that $c/\omega$ is smaller than the mean free path in the plasma. Then the plasma is concentrated on cylindrical walls, and is described by quantum fields effectively confined to those walls. Thus, the thermodynamics of these ultra-vortical plasmas may well differ very radically from those with much larger values of $c/\omega$, because the plasma becomes structured in a specific way.

These cylindrical structures could leave signatures in the electromagnetic radiation generated by the system, and these would not be greatly affected by the evolution of the plasma; this may ultimately give us a direct observational probe of these ultra-rapidly rotating systems. Thus the ideas of \cite{kn:buzztuch} may give us a way to confirm or disprove theoretical ideas regarding these plasmas, something that is currently not possible \cite{kn:hot}.

For our purposes, such extreme values of $\mathfrak{j}$ are of great interest, because the development of the structures posited in \cite{kn:buzztuch} means that, when $c/\omega$ approaches the mean free path, the relevant phase space will be severely constricted. This would lead to a reduction in the entropy density at fixed temperature, in a manner similar to the \emph{known} \cite{kn:hof,kn:bali} effects of ultra-strong magnetic fields on the phase spaces of strongly interacting matter. This may mean that ultra-rapidly rotating plasmas have a surprisingly small value of $\mathfrak{R}$ when compared with more slowly rotating plasmas at a similar temperature. That is, the increase we discussed above may be \emph{reversed} at the highest specific angular momenta.

In summary, then: on the basis of the (admittedly very limited) phenomenological evidence at our disposal, we tentatively expect large temperatures and moderately large specific angular momenta to \emph{increase} $\mathfrak{R}$ but with an upper bound; however, extremely large specific angular momenta might well eventually \emph{decrease} it, just as extremely large magnetic fields do. We hope for further clarity in the near future from the rapid rate of progress in this field.

Again, one can reasonably hope that these expectations apply to \emph{all} strongly coupled matter, not just to the QGP; and this includes the boundary ``matter'' that appears in the AdS/CFT correspondence. If we accept this, then holographic methods become available to us, and we can use them to render this qualitative analysis more quantitative.

Henceforth we use conventional, not natural or particle theory, units. See \cite{kn:110} for the motivation for this.

\addtocounter{section}{1}
\section* {\large{\textsf{4. $\mathfrak{R}$ in the Bulk}}}
We now discuss the bulk geometry relevant to computing the bulk version of $\mathfrak{R}$. We will approach this question in the most straightforward way, by studying a ``large'', singly-rotating, hot (therefore \emph{non}-extremal) AdS$_5$-Kerr black hole \cite{kn:hawk,kn:cognola,kn:gibperry} with a large specific angular momentum (angular momentum per unit mass). For this bulk geometry corresponds holographically \cite{kn:nat} to a four-dimensional field theory which might model a generic strongly-coupled plasma, of the kind generated \cite{kn:STAR,kn:franc,kn:prefut,kn:hot,kn:sign} in (normally, in reality, non-central) heavy-ion collisions\footnote{Work on using gauge-gravity duality to model the Quark-Gluon Plasma continues: see for example \cite{kn:romulo,kn:buch}. For the rotating case, see \cite{kn:bin,kn:rotating,kn:nel}.}. Generic realistic plasmas reach thermodynamic equilibrium during their brief lifetimes, have very high temperatures, and large specific angular momenta, matching the AdS$_5$-Kerr geometries of precisely this kind\footnote{We avoid the extremal case, for the obvious reason that such black holes are not hot and do not correspond to a plasma. In fact, the near-extremal case is proving to be far more complex than was previously supposed: see for example \cite{kn:hao,kn:horo1,kn:horo2,kn:horo3}.}.

That is, we compute as usual the Hawking temperature, the specific angular momentum and specific entropy of such a black hole, and assume that these quantities match their counterparts on the boundary.

We begin with a discussion of the physics of AdS$_5$-Kerr black holes. This discussion has to be more extensive than one might expect, because recently it has been pointed out that there is a small ---$\,$ but, for us, crucial ---$\,$ ambiguity in the definition of the physical mass of such black holes, and we need to consider this.

The singly-rotating AdS$_5$-Kerr metric takes the form

\begin{flalign}\label{A}
g(\textsf{AdSK}_5)\; = \; &- {\Delta_r \over \rho^2}\left[\,\m{d}t \; - \; {a \over \Xi}\,\m{sin}^2\theta \,\m{d}\phi\right]^2\;+\;{\rho^2 \over \Delta_r}\m{d}r^2\;+\;{\rho^2 \over \Delta_{\theta}}\m{d}\theta^2 \\ \notag \,\,\,\,&+\;{\m{sin}^2\theta \,\Delta_{\theta} \over \rho^2}\left[a\,\m{d}t \; - \;{r^2\,+\,a^2 \over \Xi}\,\m{d}\phi\right]^2 \;+\;r^2\cos^2\theta \,\m{d}\psi^2,
\end{flalign}
where
\begin{eqnarray}\label{B}
\rho^2& = & r^2\;+\;a^2\cos^2\theta, \nonumber\\
\Delta_r & = & \left(r^2+a^2\right)\left(1 + {r^2\over L^2}\right) - 2M,\nonumber\\
\Delta_{\theta}& = & 1 - {a^2\over L^2} \, \cos^2\theta, \nonumber\\
\Xi & = & 1 - {a^2\over L^2}.
\end{eqnarray}
Here $L$ is the asymptotic curvature length scale, $M$ and $a$ are parameters (with units of length squared and length respectively) which describe the local geometry, and the angular coordinates are Hopf coordinates on the (topological) three-sphere. The rotation is in the direction of $\phi$. Note carefully that, in contrast to the familiar asymptotically flat Kerr case, $a$ here is \emph{not} equal to the angular momentum per unit mass, though of course the two are related (see below).

In equations (\ref{A}) and (\ref{B}) we are using Boyer-Lindquist-like coordinates. Having chosen to describe the metric with these coordinates, we must \emph{consistently use them throughout} our work; we will use them to describe both the conformal infinity and the interior of the black hole.

We draw the reader's attention to the constant factor $\Xi$ which accompanies each appearance of $\phi$ : it is the cause of the ambiguity we are about to discuss (and of the extreme complexity of the functions we will encounter later), yet it has to be present. For, as is explained in detail in \cite{kn:109}, if we neglect it, then the three-dimensional subspaces of the form $t = $ constant, $r = $ constant will have conical singularities for all non-zero $a$, \emph{and so will the corresponding spatial sections at infinity}.

In a purely non-holographic context, one might actually be prepared to accept this: after all, in the ordinary asymptotically flat, four-dimensional Kerr geometry, the \emph{ergosurfaces} do in fact have conical singularities at the poles \cite{kn:kerr} for all non-zero angular momenta, yet this does not trouble us; why, then, should we be reluctant to accept (for example) an event horizon with a conical singularity? Here, however, the space at infinity is intended to represent the actual physical space on which the dual field theory propagates, and we do not want conical singularities there. This is the real reason why we are forced to include all of the factors of $\Xi$ in the bulk metric: if we wish to use holography, including them is required by the demand that the dual theory be well-behaved.

We see then that the quantity $\Xi$ has a basic role to play in the holography of such black holes. Furthermore, its presence in (\ref{A}) has major consequences: it affects the area of the event horizon, and so it appears explicitly in the expression for the entropy:
\begin{equation}\label{C}
\mathcal{S}_{\textsf{}}\; =\; {\pi^2k_{\textsf{B}}c^3\left(r_{\textsf{H}}^2 + a^2\right)r_{\textsf{H}}\over 2G_5 \hbar\,\Xi},
\end{equation}
where $k_{\textsf{B}}$ is the Boltzmann constant, $c$ is the speed of light, $G_5$ is the five-dimensional gravitational constant, $\hbar$ is the reduced Planck constant, and $r_{\textsf{H}}$ is the value of the radial coordinate at the event horizon.

This in turn means that $\Xi$ has to be involved in various other parameters describing the thermodynamics of this black hole, including the expression for the \emph{physical mass} $\mathcal{M}$ (which is \emph{not} a universal multiple of $M$ for these black holes); but the way in which it should appear is a matter of debate.

This question has recently been clarified (and, in our view, settled) by the work reported in \cite{kn:gaogao}. The basic issue is the following.

Because these black holes are not asymptotically flat, the usual (``ADM'') way of assigning a physical mass is not applicable. There is in fact a large literature on ``conserved charges in non-asymptotically-flat spacetimes'': see for example \cite{kn:hol}. There is no universally agreed way of dealing with such problems.

In particular, it is not at all clear how $\Xi$ should appear in the formula for $\mathcal{M}$, and this has led to a plethora of proposals of varying degrees of plausibility: see the detailed discussion in \cite{kn:gibperry}. The leading suggestions have however all taken the form (in the five-dimensional, singly rotating case)
\begin{equation}\label{D}
\mathcal{M}_{\textsf{}}\;=\;{\pi M c^2\left(3 - {a^2\over L^2}\right)\over 4\,G_5\,\left(1 - {a^2\over L^2}\right)^{\alpha}}
\end{equation}
where $\alpha$, which need not be an integer, is the disputed parameter. For example, $\alpha = 1$ was suggested in \cite{kn:hawk}, while \cite{kn:cognola} and \cite{kn:gibperry} argued for $\alpha = 2.$ This latter suggestion has long been generally accepted (see for example \cite{kn:mcong,kn:bay1,kn:bay2}), primarily because it leads to a simple and natural formulation of the First Law of black hole thermodynamics for these black holes. We shall refer to $\mathcal{M}$ with $\alpha = 2$ as $\mathcal{M}_2.$

Recently, however, it has been pointed out that this way of ensuring compatibility with the First Law is not unique \cite{kn:piotr,kn:gaogao}. In particular\footnote{Reference \cite{kn:piotr} is primarily concerned with the de Sitter case, while \cite{kn:gaogao} focuses on the Anti-de Sitter case of interest to us here.}, it is argued in \cite{kn:gaogao} that an alternative way of doing this leads to (\ref{D}) but with $\alpha = 3/2.$ The additional factor of $\Xi^{1/2}$ arises from the freedom to normalise the timelike Killing vector in the exterior spacetime; this freedom arises from the simple fact that one cannot in this case normalise the Killing vector by exploiting asymptotic flatness.

This particular factor arises from the argument that it is very natural to require the expression for the mass in the First Law to be an integrable differential, leading to the appropriate version of the Smarr relation. It is shown in \cite{kn:gaogao} (in two different ways) that in fact we \emph{must} use $\alpha = 3/2$ if, as is the case here, we wish to employ Boyer-Lindquist coordinates consistently\footnote{We are therefore consistently using coordinates which ``rotate at infinity''. The origin of this terminology is the fact that frame-dragging \emph{persists to infinity} for asymptotically AdS-Kerr black holes. That is, relative to an observer located at one of the poles, a zero-angular-momentum particle located on the equator of the sphere at infinity has a non-zero angular velocity $- ac/L^2$. This is just a straightforward extension of ordinary frame-dragging, which is a real and well-understood effect. One can change to coordinates which do not ``rotate at infinity'', so that this effect is concealed; this is very useful for some purposes, but the results of \cite{kn:gaogao} show that it is not necessary.}. We will call this version of the mass $\mathcal{M}_{3/2}$, and we use it exclusively. We will argue below that, in fact, this version leads to \emph{much more reasonable physics} than $\mathcal{M}_2.$

The specific entropy of the black hole is, if we use equation (\ref{D}) with $\alpha = 3/2$,
\begin{equation}\label{E}
\mathfrak{s}_{\textsf{}}\;=\;{4 \pi k_{\textsf{B}} c r_{\textsf{H}}\left(1 - {a^2\over L^2}\right)^{1/2}\over \hbar \left(3 - {a^2\over L^2}\right)\left(1 + {r_{\textsf{H}}^2\over L^2}\right)}.
\end{equation}
The angular momentum of the black hole is
\begin{equation}\label{F}
\mathcal{J}_{\textsf{}}\;=\;{\pi M c^3 a\over 2\,G_5\,\left(1 - {a^2\over L^2}\right)^2},
\end{equation}
so (again using (\ref{D})) the specific angular momentum, $\mathfrak{j}_{\textsf{}}$, is
\begin{equation}\label{G}
\mathfrak{j}_{\textsf{}}\;=\;{2 a c \over \left(3 - {a^2\over L^2}\right)\left(1 - {a^2\over L^2}\right)^{1/2}}.
\end{equation}
Notice an important point: the \emph{specific} entropy and angular momentum, unlike the entropy and angular momentum themselves, do not depend on $G_5,$ the actual value of which consequently never appears in our considerations. This is one reason why they are the parameters which should appear in a holographic model. ($r_{\textsf{H}}$ can be computed directly given \emph{only} the temperature and $\mathfrak{j}\;$: see below.)

We see that $\mathfrak{j}_{\textsf{}}$ increases monotonically from $0$ to infinity as $a$ ranges from $0$ to (nearly) $L$. \emph{This is a major difference between $\mathcal{M}_2$ and $\mathcal{M}_{3/2}.$} For in the case of $\mathcal{M}_2,$ $\mathfrak{j}_{\textsf{}}$ is also a monotonically increasing function of $a$, but it is \emph{bounded} above by $cL$ (because in that case the factor $\left(1 - {a^2\over L^2}\right)^{1/2}$ is absent from the denominator in equation (\ref{G})).

The existence of an upper bound on the specific angular momentum in terms of $L$ is not easily understood. On the boundary, $L$ does have a physical meaning \cite{kn:nat}: it is determined by the string length scale $\ell_s$ and by the 't Hooft coupling parameter $\lambda$, through the relation ${\ell_s^4\over L^4}\;=\;{1\over \lambda}.$ It is very far from clear however why $\ell_s$ and $\lambda$ should play a role in constraining $\mathfrak{j}.$ The choice of $\mathcal{M}_{3/2}$ relieves us of this apparently superfluous restriction, and we consider this to be conclusive evidence that $\mathcal{M}_{3/2}$ must be used instead of $\mathcal{M}_2$ (if we consistently use Boyer-Lindquist coordinates).

The parameter $a$ is geometric rather than physical, in the sense that it is the parameter that occurs explicitly in the metric tensor. Inverting equation (\ref{G}) allows us, however, to regard it as a (complicated) monotonically increasing function of $\mathfrak{j},$ and that is how we will interpret it henceforth.

The Hawking temperature of this black hole is (combining \cite{kn:gibperry,kn:gaogao})
\begin{equation}\label{H}
T_{\textsf{}}\;=\;{\hbar c r_{\textsf{H}}\over 2 \pi k_{\textsf{B}}\left(1 - {a^2\over L^2}\right)^{1/2}}\left[{1 + {r_{\textsf{H}}^2\over L^2}\over r_{\textsf{H}}^2 + a^2} \,+\, {1\over L^2}\right].
\end{equation}

We propose to make use of this relation in an unusual way. Solving (\ref{H}) for $r_{\textsf{H}}$, we regard it as a known function of $T_{\textsf{}}$ and $a$ (therefore, in view of our discussion of $a$ above, of $T_{\textsf{}}$ and $\mathfrak{j}_{\textsf{}}$). Then equation (\ref{E}) gives the specific entropy $\mathfrak{s}_{\textsf{}}$ as a function of $T_{\textsf{}}$ and $\mathfrak{j}_{\textsf{}}$ (and $\hbar$ and $k_{\textsf{B}}$, but not $G_5$). Thus we can express $\mathfrak{s}\,T/\hbar$ as such a function, and now we can use holography to study the dependence of $\mathfrak{R}$ as a function of $T_{\textsf{}}$ and $\mathfrak{j}_{\textsf{}}$.

Before proceeding to that, let us recall \cite{kn:edwit} that, when $a = 0$, there are in fact no real solutions for $r_{\textsf{H}}$ if the temperature falls below a certain value, $\sqrt{2}\hbar c / \left(\pi k_{\textsf{B}}L\right)$. Above that temperature, there are two (or three) solutions. The largest defines a ``large'' black hole, which describes an equilibrium state for the boundary theory \cite{kn:ruong}. Below that temperature, the holographic model predicts that strongly coupled matter at equilibrium simply does not exist. Thus, in evaluating the black hole radius, we always choose the largest solution at $a = 0$, following this branch as $a$ is increased. (That is, for us, a ``large'' AdS$_5$-Kerr black hole is the one corresponding to the largest value of $r_{\textsf{H}}$ for a given value of $a$.)

For our later purposes, it will be necessary to have an explicit understanding of the ways in which $r_{\textsf{H}}$ varies as a function of $T_{\textsf{}}$ and $a$.

As $a$ increases at some fixed value of $T$, $r_{\textsf{H}}$ is a  \emph{decreasing} function of $a/L$ (see Figure 1). This is not remarkable: it is in fact precisely what happens when the specific angular momentum of an asymptotically flat Kerr black hole of fixed temperature is increased \cite{kn:109}. In that case there is a non-zero lower bound on the radius. Here, however, that is not so: as $\mathfrak{j}$ tends to infinity, so that $a/L$ tends to unity, the radius actually approaches zero, as one sees directly from equation (\ref{H}).

\begin{figure}[!h]
\centering
\includegraphics[width=0.75\textwidth]{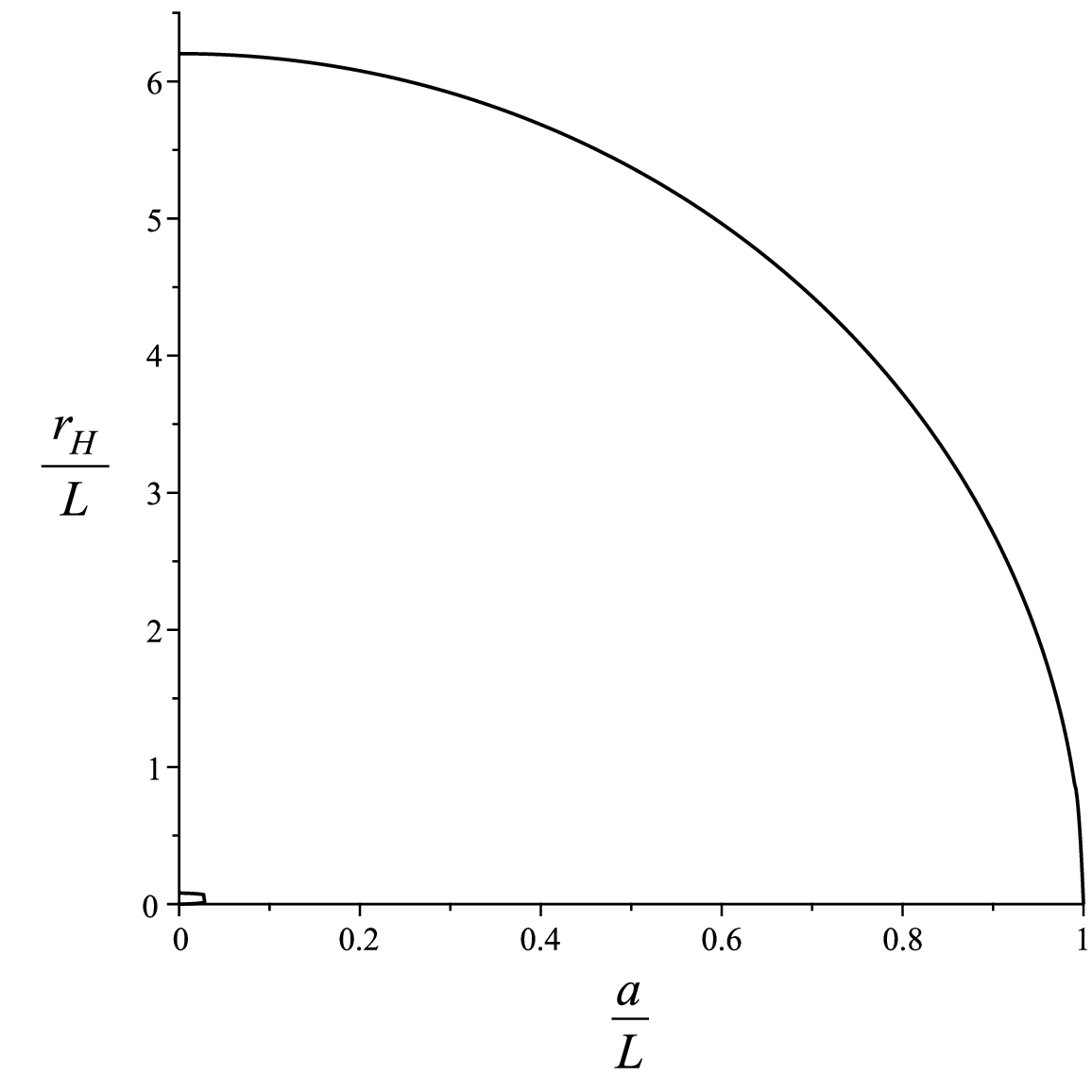}
\caption{Scaled event horizon radius for ``large'' AdS$_5$-Kerr with $T^*L = 2$, as a function of $a/L$. Here $T^* = k_{\textsf{B}}T/(\hbar c)$.}
\end{figure}

As $T$ increases at a fixed value of $a$ or $\mathfrak{j}$, by contrast, the event horizon radius (for the ``large'' black holes of interest to us here) increases monotonically: see Figure 2. 

\begin{figure}[!h]
\centering
\includegraphics[width=0.80\textwidth]{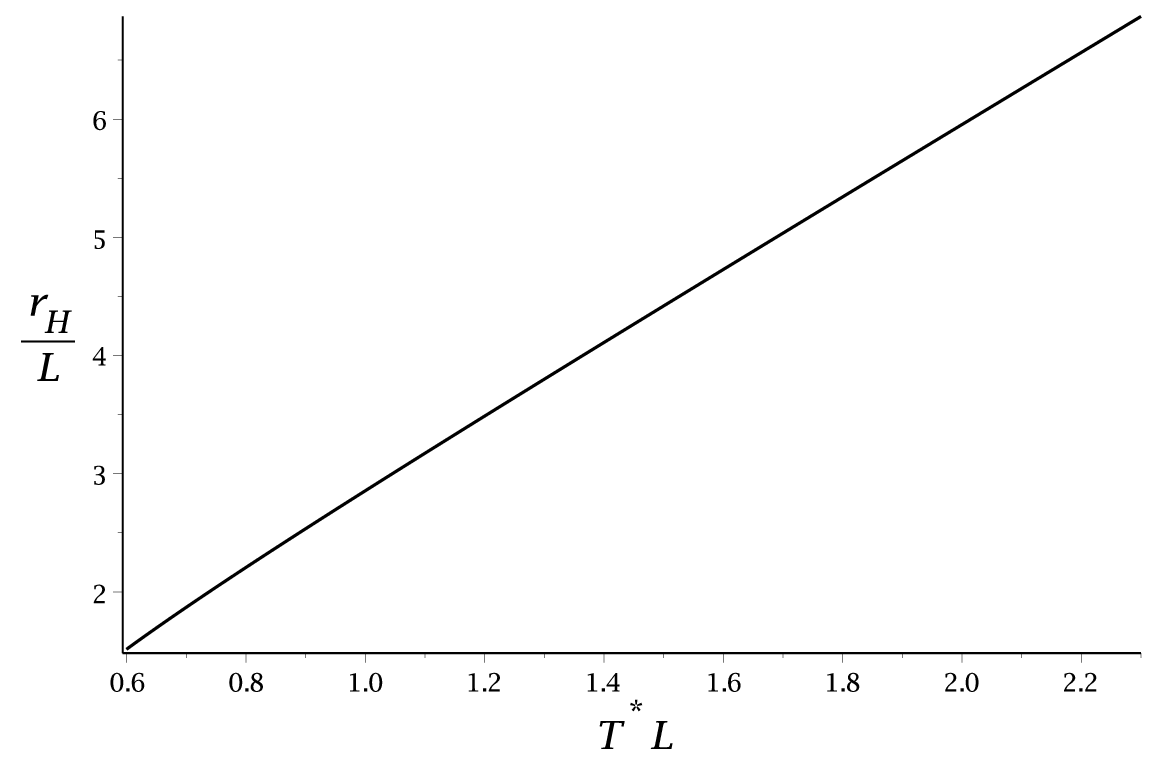}
\caption{Scaled event horizon radius for ``large'' AdS$_5$-Kerr with $\mathfrak{j}/(cL) = 0.2 $, as a function of $T^*L$. Here $T^* = k_{\textsf{B}}T/(\hbar c)$.}
\end{figure}

We note in passing that, just as $a$ is fixed by $\mathfrak{j}$, the other geometric parameter in the metric, $M$, can now also be regarded as a known function of $T_{\textsf{}}$ and $\mathfrak{j}_{\textsf{}}$, by using the definition of $r_{\textsf{H}}$, namely
\begin{equation}\label{I}
M\;=\; {1\over 2}\left(r_{\textsf{H}}^2+a^2\right)\left(1 + {r_{\textsf{H}}^2\over L^2}\right).
\end{equation}
Thus, the bulk geometry is explicitly fixed holographically in terms of the temperature and specific angular momentum of the boundary matter.

Notice from equation (\ref{I}) that, since the radius is an increasing function of the temperature for fixed angular momentum, and the physical mass is a multiple of $M$, then, all else being equal, more massive ``large'' black holes of this kind have higher temperatures: that is, the specific heat is positive.

This completes our study of the physics of AdS$_5$-Kerr black holes and of the physical interpretation of their parameters. Now we turn to the study of $\mathfrak{s}\,T/\hbar$ as a function of impact energy and impact parameter, that is, of $T$ and $\mathfrak{j}$.

\addtocounter{section}{1}
\section* {\large{\textsf{5. $\mathfrak{s}\,T/\hbar$ for AdS$_5$-Kerr as a Function of $T$ and $\mathfrak{j}.$}}}

We now compute $\mathfrak{s}\,T/\hbar$ for the AdS$_5$-Kerr black hole.

From equation (\ref{E}) we now have
\begin{equation}\label{J}
\mathfrak{R}\; = \; {4\pi k_{\textsf{B}} c T_{\textsf{}} r_{\textsf{H}}\left(1 - {a^2\over L^2}\right)^{1/2}\over \hbar^2 \left(3 - {a^2\over L^2}\right)\left(1 + {r_{\textsf{H}}^2\over L^2}\right)}.
\end{equation}

It is convenient to write this in the following form:
\begin{equation}\label{K}
\mathfrak{R} \; = \; {c^2\over \hbar}\,\Gamma\left(T^*L,\;\mathfrak{j}/(cL)\right),
\end{equation}
where $T^* = k_{\textsf{B}}T/(\hbar c)$ has units of inverse length, and where
\begin{equation}\label{L}
\Gamma\left(T^*L,\;\mathfrak{j}/(cL)\right) \;=\; {4 \pi T^* r_{\textsf{H}}\left(1 - {a^2\over L^2}\right)^{1/2}\over \left(3 - {a^2\over L^2}\right)\left(1 + {r_{\textsf{H}}^2\over L^2}\right)}.
\end{equation}
Note that $\Gamma,\; T^*L,$ and $\mathfrak{j}/(cL)$ are all dimensionless (both in conventional and in natural units). In fact, $\Gamma$ corresponds to the quantity listed in the Table given earlier.

\begin{figure}[!h]
\centering
\includegraphics[width=0.8\textwidth]{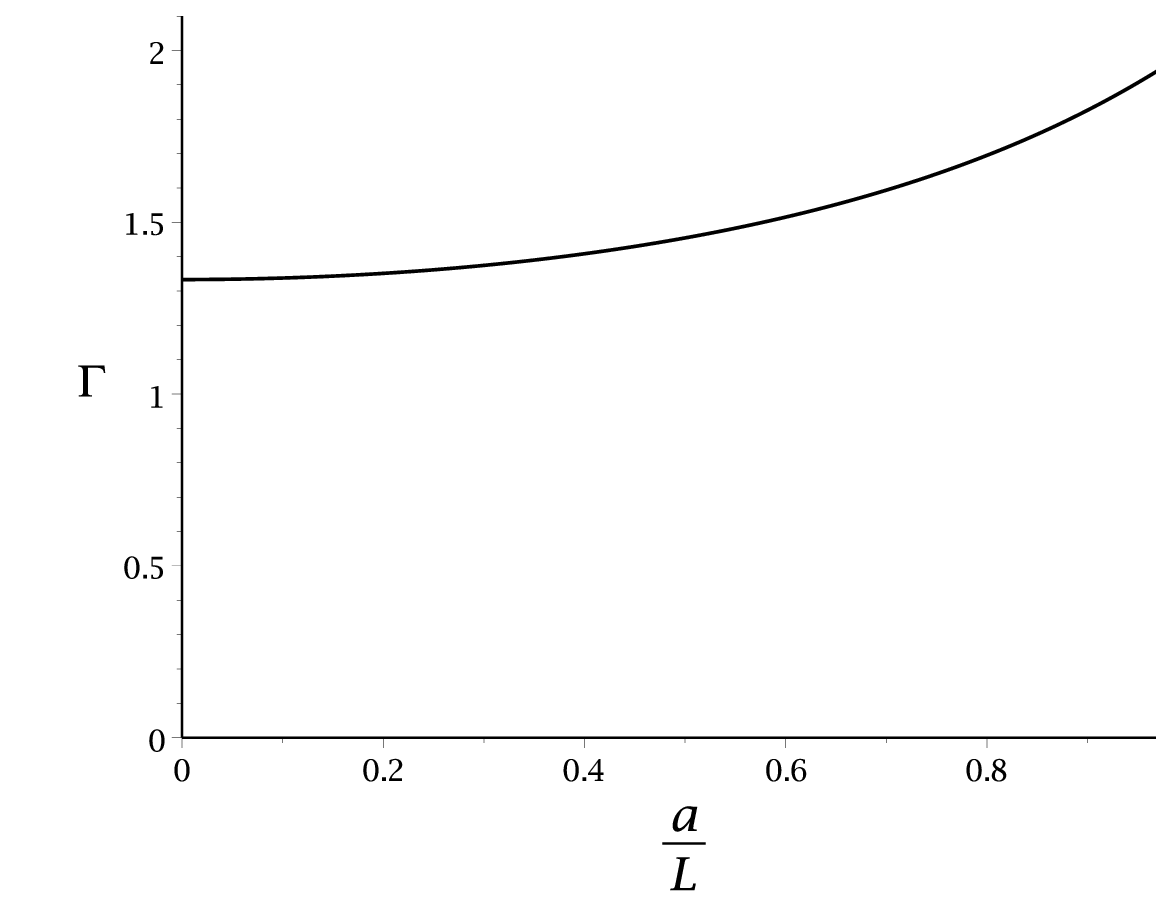}
\caption{ $\Gamma\left(T^*L,\;\mathfrak{j}/(cL)\right)$ for AdS$_5$-Kerr, as a function of $a/L$ (that is, indirectly as a function of $\mathfrak{j}/(cL)$), when $T^*L$ is taken for illustrative purposes to be equal to 50. As $a/L$ increases, the function approaches a value just under 2 from below (but it then suddenly falls towards zero).}
\end{figure}

Recalling that $a$ is a known function of $\mathfrak{j}$, and that $r_{\textsf{H}}$ is (through equation (\ref{H})) a function of $T$ and $a$, we think of $\Gamma$ as a function of two variables, $T$ and $\mathfrak{j}$. It turns out to be an extremely complicated function of its those variables, and to determine its detailed properties requires a numerical treatment.

One can however show analytically  that $\Gamma,$ interpreted in this way, is a \emph{bounded} function, and this bound can be evaluated exactly (with assistance from MAPLE$^{\textsf{TM}}$). One has, for fixed $\mathfrak{j}$, that is, fixed $a$, 
\begin{equation}\label{ZZZZZZ}
\lim_{T \rightarrow \infty}\Gamma \;=\; {4\over 3 - {a^2\over L^2}}.
\end{equation}
In particular, for central collisions ($a = 0$) one sees that the bound is $4/3$, that is, \emph{precisely the value we obtained in Section 3}, on general thermodynamic grounds, for the value of $\mathfrak{s}\,T$ (in natural units) in central collisions at the highest impact energies \cite{kn:sign}. It is remarkable and encouraging that this simple holographic model is able to reproduce the thermodynamic result exactly.

One sees that the bound increases steadily with $a$: angular momentum permits, at sufficiently large temperatures, values of $\Gamma$ beyond $4/3$. As $\mathfrak{j}$ tends to infinity (that is, as $a/L$ tends to unity), the upper bound tends to $2$. To be precise, we have $\Gamma < 2$ and
\begin{equation}\label{LLL}
\lim_{\mathfrak{j} \rightarrow \infty}\;\lim_{T \rightarrow \infty}\Gamma \;=\; 2.
\end{equation}
To be still more precise: for each fixed value of $T^*L,$ $\Gamma$, regarded as a function of $\mathfrak{j}/(cL)$, has a global maximum, which occurs at ever-increasing values of $\mathfrak{j}/(cL)$ as $T^*L$ is adjusted upwards. This maximum tends upwards to 2 as $T^*L$ is taken to infinity. See for example Figure 3, where $T^*L$ has been fixed at a large value, 50.

We see that holography predicts the existence of a \emph{universal} upper bound on $\mathfrak{R}$ for strongly coupled matter:
\begin{equation}\label{LL}
\mathfrak{R} \; < \; {2 c^2\over \hbar} \; \approx \; 1.70 \times 10^{51}/(\m{kg}\cdot \m{s}).
\end{equation}
This is a very large value by the standards of ordinary matter; nevertheless, we saw earlier that a phenomenological model of heavy-ion collisions at moderate impact energies and low centralities predicts values for $\mathfrak{s}\,T/\hbar$ in the QGP which are below this bound but actually not far below it (ranging from about $1 \times c^2/\hbar$ to around $1.3 \times c^2/\hbar$). More interestingly still, the theory clearly indicates that large angular momenta might make it possible to attain values for $\mathfrak{R}$ which are greater than $4/3$ and thus thermodynamically forbidden for non-rotating strongly coupled matter.

We now study how $\Gamma$ varies as $T$ and $\mathfrak{j}$ are varied independently. We start with $T$.

\addtocounter{section}{1}
\subsection* {\large{\textsf{5.1 $\;\;\;\mathfrak{s}\,T/\hbar$ for AdS$_5$-Kerr as a Function of $T$}}}

In order to compare the values of $\Gamma$ predicted by the holographic model with the values in the Table given above, we fix $\mathfrak{j}/(cL)$ at a relatively small value, $0.2$ (because \cite{kn:sahoo} only considers low-centrality collisions; this corresponds to $a/L \approx 0.2804$). Substituting this into equation (\ref{L}) in the manner explained above, we obtain a function of $T^*L,$ as shown in Figure 4.

\begin{figure}[!h]
\centering
\includegraphics[width=0.99\textwidth]{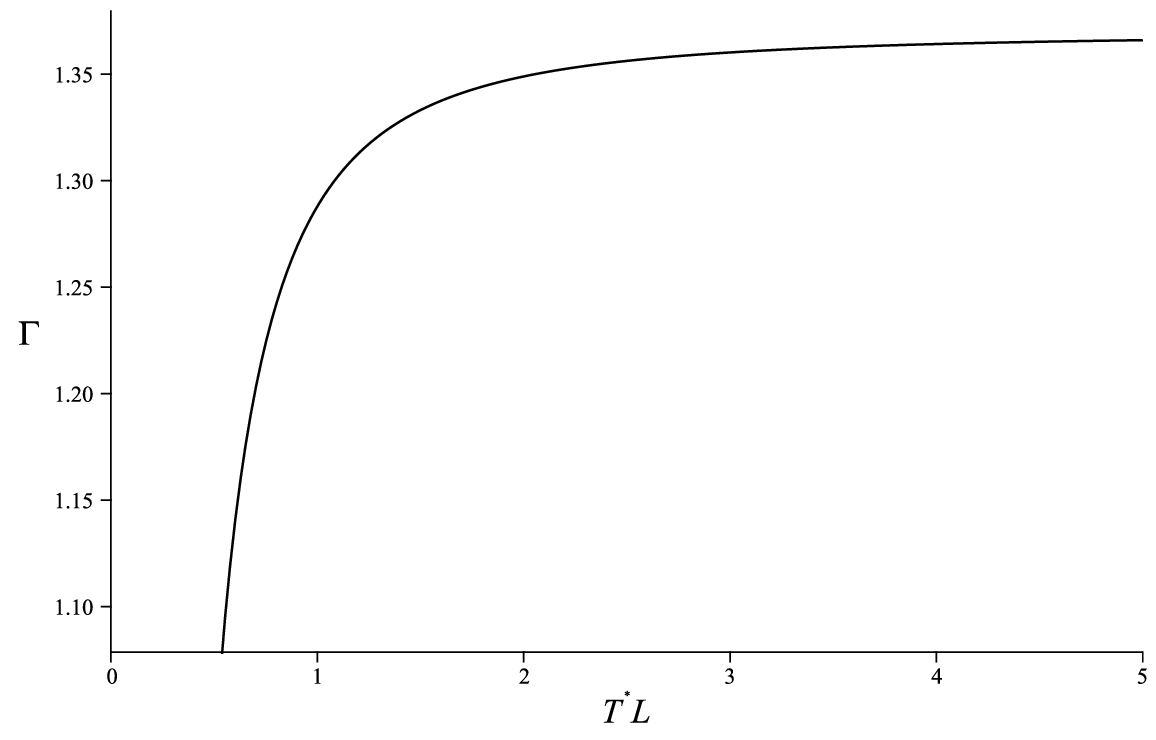}
\caption{ $\Gamma\left(T^*L,\;\mathfrak{j}/(cL)\right)$ for AdS$_5$-Kerr, as a function of $T^*L$, when the specific angular momentum is small, $\mathfrak{j}/(cL) = 0.2$.}
\end{figure}

The predicted values for $\Gamma$ grow monotonically, but are bounded above by a value slightly larger than $4/3$ at arbitrarily high impact energies, as we discussed earlier. 

We also saw earlier that ``large'' AdS$_5$ black holes have temperatures which are necessarily bounded below (by $\sqrt{2}\hbar c / \left(\pi k_{\textsf{B}}L\right)$), and this implies that the holographic model puts a lower bound on $\Gamma$. This lower bound, when $a \approx  0.2804$ as in Figure 4, is 
\begin{equation}\label{YYYYY}
\Gamma \;>\; \approx 0.922.
\end{equation}
This too agrees remarkably well with the phenomenological model \cite{kn:sahoo} discussed in Section 3, where we saw that, because the temperature of the QGP is bounded below, values of $\Gamma$ below about $1$ are not to be expected (in central collisions). 

On the other hand, note that the horizontal axis in Figure 4 involves $L$, the precise value of which is not known. Caution is therefore called for. Nevertheless we claim that the holographic model reproduces the physical data unexpectedly well.

We have seen that angular momentum permits values of $\Gamma$ which are higher than those allowed in central collisions. One naturally asks whether it might also permit values lower than about $1$. We will now investigate this.

\addtocounter{section}{1}
\subsection* {\large{\textsf{5.2 $\;\;\;\mathfrak{s}\,T/\hbar$ for AdS$_5$-Kerr as a Function of $\mathfrak{j}$}}}

We now ask how $\Gamma\left(T^*L,\;\mathfrak{j}/(cL)\right)$ behaves when we fix the temperature and let $\mathfrak{j}$ vary.

Figure 3 shows this function when $T^*L = 2,$ the value being chosen for convenience. Other values of $T^*L$ above the lowest permitted value yield graphs of the same general form, though the maximum becomes more sharply peaked at high temperatures: see for example Figure 3. It is convenient to use $a/L$ on the horizontal axis, since doing so allows us to survey the full range of possible values for $\mathfrak{j}$, including extremely large ones: recall that $a/L$ tends to unity, as a monotonically increasing function of $\mathfrak{j},$ as the latter tends to infinity.

\begin{figure}[!h]
\centering
\includegraphics[width=0.70\textwidth]{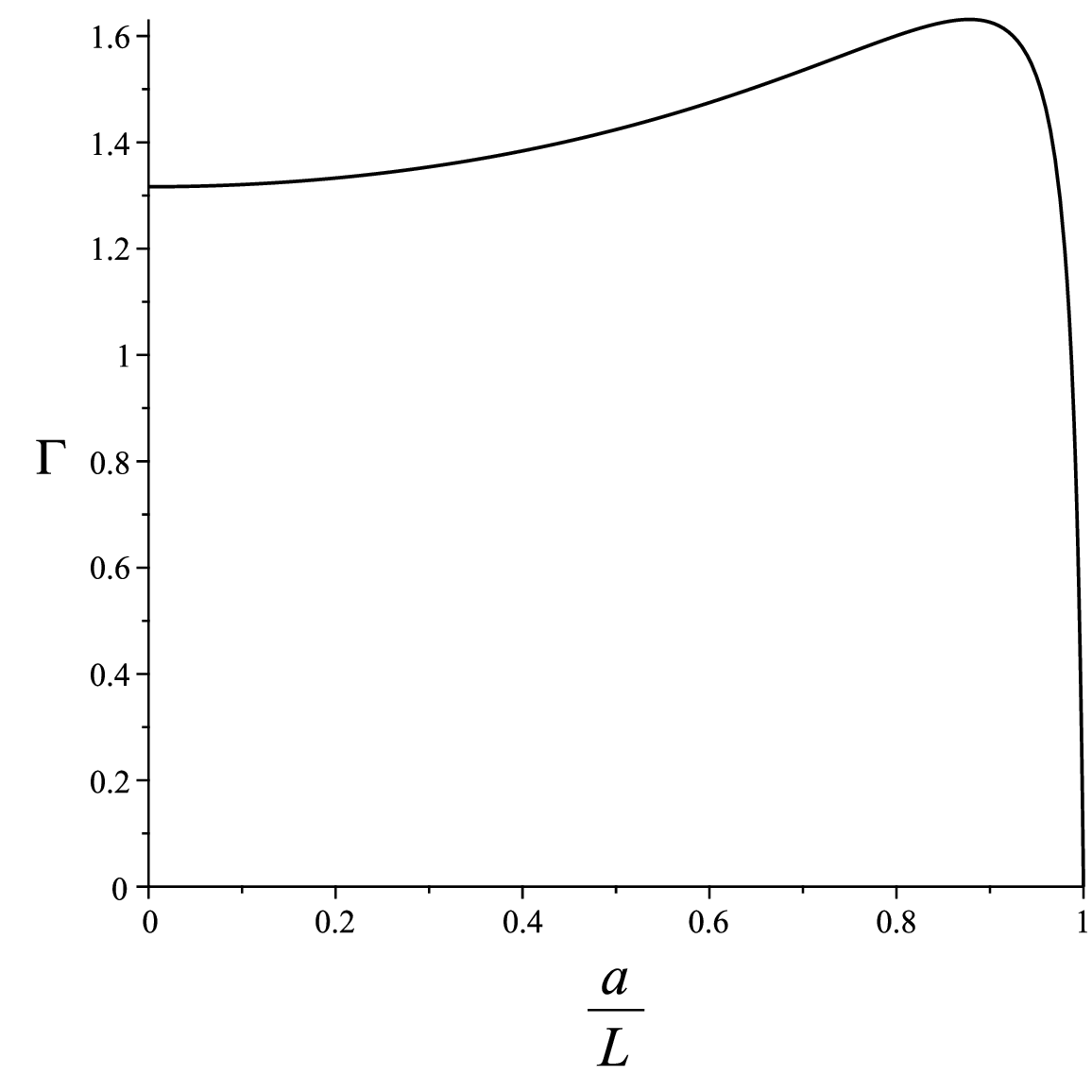}
\caption{$\Gamma\left(T^*L,\;\mathfrak{j}/(cL)\right)$ for AdS$_5$-Kerr, as a function of $a/L$ when $T^*L = 2.$}
\end{figure}

We see that, until $a/L$ becomes close to unity, we have here a slowly increasing function. We argued earlier that this is precisely what happens in the phenomenological model presented in \cite{kn:kshitish}. It is too early to make a definite statement here: but we can say that the (admittedly very preliminary) evidence we have does suggest that the holographic model agrees with what is predicted.

On the other hand, the work of Buzzegoli and Tuchin \cite{kn:buzztuch} led us to predict that $\mathfrak{R}$ should be suppressed at extremely high specific angular momenta, once the peak is passed. For the temperature used to construct Figure 5, the maximum is attained at around $a/L = 0.87,$ which corresponds to $\mathfrak{j} \approx 1.65\,cL$. Unfortunately, it is not possible to use the model to predict the value of $\mathfrak{j}$ at which $\Gamma$ begins to decrease, because we do not know the numerical value of $L$. However, in holography \cite{kn:nat}, $L$ is always taken to be ``large'' in the sense of being larger than any other length scale present. Thus it makes sense to say that values of $\mathfrak{j}$ beyond $cL$ are ``large'' in the holographic picture, and so $\mathfrak{j} \approx 1.65\,cL$ means that the specific angular momentum is large in this holographic sense.

Of course, we expect this unusual suppression effect to occur \emph{only} at extremely high specific angular momenta, when the angular momentum crosses some threshold, perhaps defined by the reciprocal of the mean free path. Again, this is just what one sees in Figure 5.

Clearly, the holographic model predicts that sufficiently large specific angular momenta reduce $\mathfrak{R}$, and this to \emph{any} desired degree. That is, to the extent that $\mathfrak{R}$ is a good measure of the post-equilibrium evolution of the plasma, sufficiently large specific angular momenta can (almost) bring that evolution to a complete halt. (However, we will argue later that this is in fact only true at relatively early times.) Note in this connection that, in the interpretation where $\mathfrak{R}$ is the rate of change of the ``specific complexity'', this slowdown is also expected to happen, \emph{but} only at extremely late times \cite{kn:tallarita}. Thus, the holographic model implies that ultra-high specific angular momenta give us a preview of the otherwise inaccessible remote future of the complexification process, in that particular interpretation.

To summarize: the quantity $\mathfrak{s}\,T/\hbar$, evaluated on the external spacetime of an AdS$_5$-Kerr black hole, is dual to a quantity $\mathfrak{R}$ on the boundary. What little is known about $\mathfrak{R}$ in the case of the QGP suggests that it responds in a particular way to variations of the QGP specific angular momentum and temperature. We have adduced evidence that the black hole quantity responds in a very similar way to variations of the quantities dual to $\mathfrak{j}$ and $T$.

The key point now is this: when we vary the exterior geometry of the black hole, \emph{we also inevitably vary its internal geometry} in a similar way (assuming ``no drama'' at the event horizon): in particular, such variations will affect the rate at which the internal spatial geometry evolves, because that rate is a characteristic of the interior. Now we have seen that $\mathfrak{s}\,T/\hbar$, evaluated on the exterior bulk black hole, behaves (with respect to variations of $T$ and $\mathfrak{j}$) in much the same way as $\mathfrak{R}$ on the boundary. We need to verify that these variations of $\mathfrak{s}\,T/\hbar$ also cause the rate of interior evolution to vary in a similar way. If we can do this, then we have good evidence for the existence of a duality between $\mathfrak{R}$ and the time dependence of the interior geometry, with the black hole exterior playing the role of intermediary. Even if this only works explicitly at relatively early times, we still have a good basis for an extrapolation to the entire interior of the black hole.

To be specific, we need to do the following. We need to show, first, that the rate of evolution of the interior, just under the event horizon, should always \emph{increase} as the Hawking temperature increases, with fixed specific angular momentum $\mathfrak{j}$; but it should be bounded, in fact, asymptotic to a definite finite value. Secondly, we need to show that the interior rate of evolution at fixed Hawking temperature \emph{increases} with $\mathfrak{j}$ for moderate values of $\mathfrak{j}:$ see Figure 5. At the highest values of $\mathfrak{j}$, however, it should \emph{decrease}, tending ultimately to zero, again as in Figure 5.

Let us now attempt to investigate this.

\addtocounter{section}{1}
\section* {\large{\textsf{6. Dynamics Just Under the Event Horizon}}}
There is no doubt that the spacetime just inside any event horizon is time-dependent \emph{in some sense}, simply because there is no timelike Killing vector in that part of the interior\footnote{Actually, in the specific case of the AdS$_5$-Kerr black hole, there is no timelike Killing vector anywhere in the interior, because there is no inner horizon: see equation (\ref{I}).}. Unfortunately, quantifying this time dependence is difficult. For example, as is well known in studies of numerical relativity \cite{kn:alcu}, it is possible to foliate the interior by slices that behave in many different prescribed ways as the geometry evolves. However, this is evidently a (very useful) mathematical subterfuge, usually designed to prevent the singularity from terminating the numerical evolution prematurely. The problem here is to find a foliation which is privileged, not in the sense of being ``more physical'' than any other \cite{kn:ellis}, but simply in the sense that the behaviour of the slices most effectively allows us to exhibit and quantify the fact that the geometry is not static.

We will follow, for want of anything better, the custom of attempting to measure the rate of change of the internal geometry by using the \emph{volumes} of spacelike slices. As explained earlier, however, our procedure differs in that we only use internal foliations which remain inside the event horizon.

No such programme can afford to ignore the fact that the full geometry deep inside (that is, far to the future of) the event horizon of a rotating black hole is not fully understood. However, assuming as ever that there is no ``drama'' at the event horizon, the metric given in (\ref{A}) and (\ref{B}) presumably remains at least approximately valid on spatial sections \emph{just under} the event horizon. We will focus all of our attention on this region, and abstain (until the Conclusion) from any speculation as to precisely what happens at later times.

As the event horizon is specified as a value of the radial coordinate, the obvious and natural way to study the interior time-dependence is to choose the spacelike surfaces\footnote{We may note in passing that these $r = $ constant surfaces are also distinguished by being highly symmetric, in the sense that they contain the orbits of the three spacelike Killing vectors, $\partial_t, \, \partial_{\phi},$ and $\partial_{\psi}.$} of the form $r = $ constant $\equiv r_{<\;\textsf{H}}$, where $r$ is the coordinate in (\ref{A}) and (\ref{B}); it is of course timelike in the interior (note however that it \emph{decreases} towards the future). The constant $r_{<\;\textsf{H}}$ will be some value of $r$ slightly smaller than the event horizon radius, $r_{\textsf{H}}$, in keeping with our resolve to remain just under the horizon. The reader should bear in mind that $r = 0$ is not a singular locus here except (exactly) on the equator; so even if $r_{\textsf{H}}$ and therefore $r_{<\;\textsf{H}}$ are small, which can happen here (see Figure 1), it does not necessarily follow that we are ``close'' to the singularity.

Any other interior foliation consisting of slices which remain close to the horizon will resemble this one to some extent. We can hope that picking another such foliation will not change our conclusions drastically. In short, in the absence of a canonical measure of the rate of evolution of the interior, our discussion is necessarily preliminary and our findings are tentative.

The $r = r_{<\;\textsf{H}}$  sections are infinite in the $t$ direction and for this reason (only) their volumes are infinite. However, for us, the total volume of a slice is of little interest: the volume is a means to an end, its rate of change being a measure of the fact that the interior is dynamic. In fact, even aside from this, we do not care that the volume is changing rapidly if this occurs simply because it is already very large. Our concern should be with the rate of change of the volume \emph{relative} to its current value. This is in fact very much reminiscent of our definition of $\mathfrak{R}$ as an intensive quantity: that is, as a rate which ``takes into account the resources available''.

We therefore artificially truncate $t$ to some finite range of length $\tau$, and then use the \emph{logarithmic} derivative (with respect to $r$) of the volume of a section of the form $r = $ constant. The volume turns out \cite{kn:110} to be (assuming $a \neq 0$)
\begin{equation}\label{M}
\Omega \;=\; {4\,\pi^2 \tau r\,\sqrt{|\Delta_r|}\,\left[\left(a^2 + r^2\right)^{3/2}\,-\,r^3\right]\over 3 a^2\,\Xi}.
\end{equation}
Eliminating $M$ using equation (\ref{I}), we have
\begin{equation}\label{N}
|\Delta_r|\;=\; \left(r_{\textsf{H}}^2+a^2\right)\left(1 + {r_{\textsf{H}}^2\over L^2}\right)\;-\; \left(r^2+a^2\right)\left(1 + {r^2\over L^2}\right).
\end{equation}
Thus we can express the logarithmic derivative $\lambda(\Omega)$ of $\Omega$ with respect to $r$ in terms of $r_{\textsf{H}}$, $a$, $L$, and $r$. (Of course, $\tau$ is cancelled.) It will in fact be convenient to use the non-negative dimensionless quantity $|\lambda(\Omega)|L$ instead.

We now use (\ref{G}) to express $a$ in terms of $\mathfrak{j},$ and (\ref{H}) to express $r_{\textsf{H}}$ in terms of $T$ and $\mathfrak{j}$, so that now $|\lambda(\Omega)|L$ can be regarded as a function of $T$, $\mathfrak{j}$, $L$, and $r$. It is an extremely complicated function of these variables, so a numerical investigation is required to produce comprehensible results.

In principle, we just need to fix $r$ at some value $r_{<\;\textsf{H}}$ slightly smaller than $r_{\textsf{H}}$, and then study $|\lambda(\Omega)|$ as a function of $\mathfrak{j}$ and $T$. What we are attempting to establish here is some kind of holographic relation of the schematic form
\begin{equation}\label{O}
|\lambda(\Omega)|\big(T^*L,\;\mathfrak{j}/(cL)\big)L \;\sim\; \Gamma\big(T^*L,\;\mathfrak{j}/(cL)\big) \;=\; {\hbar\over c^2}\,\mathfrak{R},
\end{equation}
(see equations (\ref{K}) and (\ref{L}) above). That is, we hope to identify the two quantities, when they can both be regarded as functions of the same variables, $T$ and $\mathfrak{j}$, evaluated on the black hole exterior (but holographically equivalent to the corresponding quantities describing the boundary matter). Here the notation $\sim$ reminds us that we do not fully understand what $\mathfrak{R}$ measures, so we are no doubt missing some (dimensionless) factor, which however we hope to be of order unity \cite{kn:tallarita}. (It also reminds us that while $r_{<\;\textsf{H}}$ is not an independent variable, it is only approximately determined by its nearness to $r_{\textsf{H}}$.)

Our plan is not to ``prove'' equation (\ref{O}), but rather to render it plausible by showing that both sides behave in much the same way as the underlying physical parameters $T$ and $\mathfrak{j}$ are varied.

Unfortunately there is an apparently simple, but in fact technically very serious problem here: $r_{\textsf{H}}$ itself depends on these parameters, as one sees in Figures 1 and 2. If $r_{<\;\textsf{H}}$ is initially slightly smaller than $r_{\textsf{H}}$, even a small increase in $\mathfrak{j}$, or a small decrease of $T$, may take us outside the event horizon and invalidate the computation of $|\lambda(\Omega)|L$ (because $t$ and $r$ become null as the event horizon is crossed, resulting in a divergence). Again, $r_{\textsf{H}}$ may increase so that $r_{<\;\textsf{H}}$ is no longer ``a little'' smaller: it may become much smaller. (However, as this does not cause a divergence, we will allow more leeway with this, for illustrative clarity; but the reader is urged to be cautious in interpreting the figures below when $r_{\textsf{H}}$ is significantly larger than $r_{<\;\textsf{H}}$.)

Here we will circumvent this problem in the most straightforward way: fixing $r$ at some value $r_{<\;\textsf{H}}$ slightly smaller than $r_{\textsf{H}}$, we will only allow $T$ and $\mathfrak{j}$ to vary within a narrow range, so that $r_{\textsf{H}}$ does not move very much. As we are not demanding precise equality in equation (\ref{O}), but rather simple qualitative agreement, we argue that this is sufficient. To put it another way, we are just sampling the parameter space to check for any blatant contradiction of (\ref{O}).

An extensive computer search has revealed no such contradictions. We present some examples.

First, let us fix $\mathfrak{j}/(cL),$ as in Figure 4, at 0.2, and pick $r_{<\;\textsf{H}} = 1$. Figure 2 shows us that, if $T^*$ is at least around $0.6/L,$ and is increased under these restrictions, the event horizon is outside the chosen slice, and moves outwards, and so there should be no difficulty (though, as explained, one should be a little cautious about very high temperatures, since then $r_{<\;\textsf{H}} = 1$ puts us deep inside). The graph of $|\lambda(\Omega)|\big(T^*L,\;\mathfrak{j}/(cL) = 0.2\big)L$ on this domain is shown as Figure 6.

\begin{figure}[!h]
\centering
\includegraphics[width=0.80\textwidth]{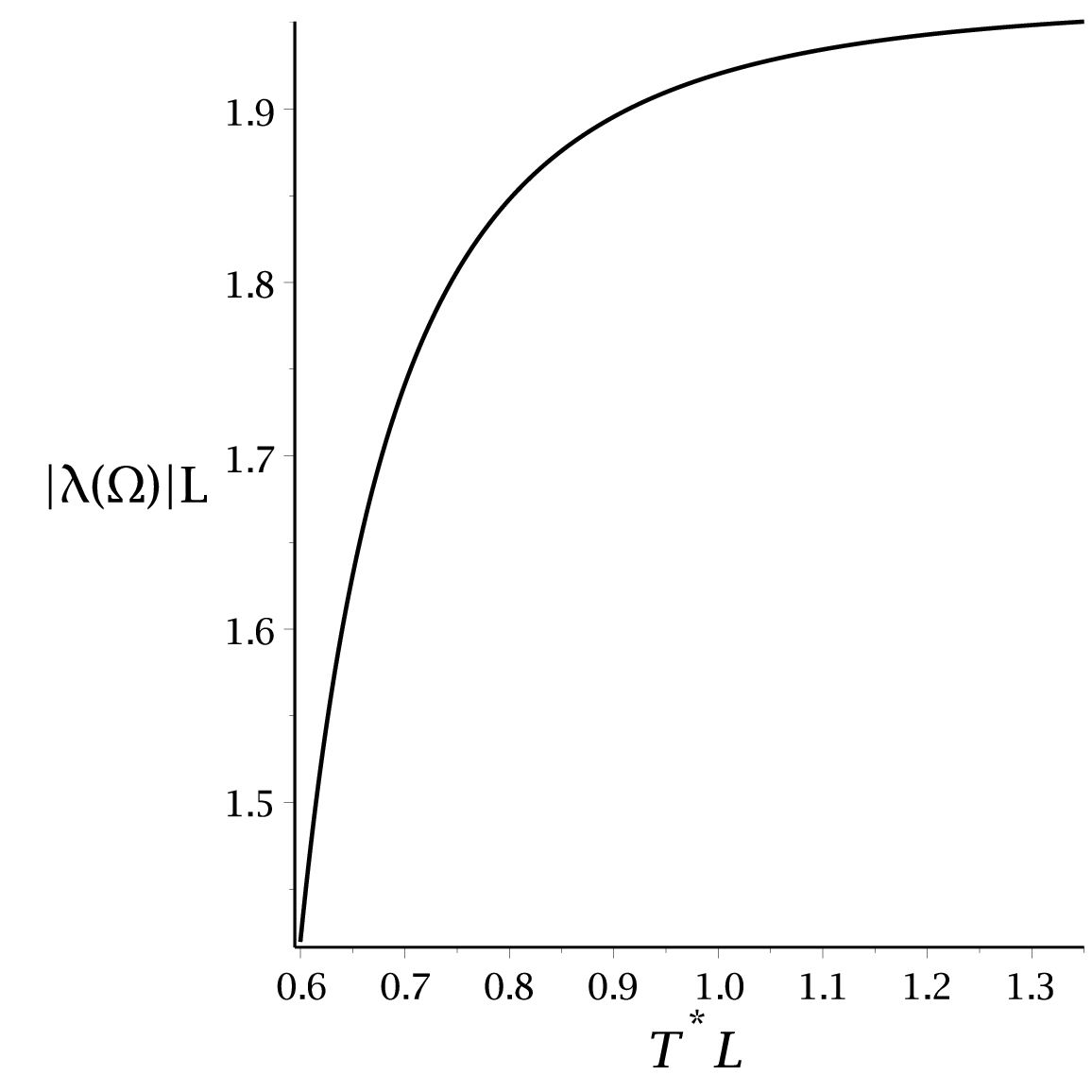}
\caption{ $|\lambda(\Omega)|\big(T^*L,\;\mathfrak{j}/(cL)\big)L$ for AdS$_5$-Kerr, as a function of $T^*L$, when the specific angular momentum is small, $\mathfrak{j}/(cL) = 0.2$, and $r_{<\;\textsf{H}} = 1$.}
\end{figure}

This is to be compared with Figure 4. The graphs are broadly similar, portraying functions which increase but can be expected to be asymptotic to a constant (this would be clearer in Figure 6 if we allowed higher temperatures; using MAPLE$^{\textsf{TM}}$ one can show that, with these parameter values, $\lim_{T \rightarrow \infty}|\lambda(\Omega)|\big(T^*L,\;\mathfrak{j}/(cL) = 0.2\big)L \approx 1.962$, so this is the upper bound in this case). That is, the rate of evolution of the volume of a slice is always increased by increasing the Hawking temperature, but there is an upper bound to this increase, just as happens to the claimed dual quantities outside the black hole and at the boundary of the spacetime.

If we now do the opposite and fix the temperature, while allowing $\mathfrak{j}/(cL)$ to increase, we see from Figure 1 that we have a much more difficult problem to handle, because $r_{\textsf{H}}$ immediately drops. (As usual, we use $a/L$ as a proxy for $\mathfrak{j}/(cL)$: recall that the former tends monotonically to unity as the latter tends to infinity.) The decline is gentle for low values of $\mathfrak{j}/(cL)$, but precipitous for large values.

From Figure 1 we see that, with $T^*L = 2$ (as in Figure 5), $r_{<\;\textsf{H}} = 5.5L$ will put us just under the event horizon for $a/L$ in the range from about 0.25 to 0.35, so let us take our sample there. The result is shown as Figure 7, to be compared with the same range of $a/L$ values in Figure 5. The shapes of the two graphs agree; at fixed temperature, and small specific angular momenta, an increase in the latter causes the rate of change of the interior geometry to increase, just as expected if there is a holographic correspondence in this case.

\begin{figure}[!h]
\centering
\includegraphics[width=0.80\textwidth]{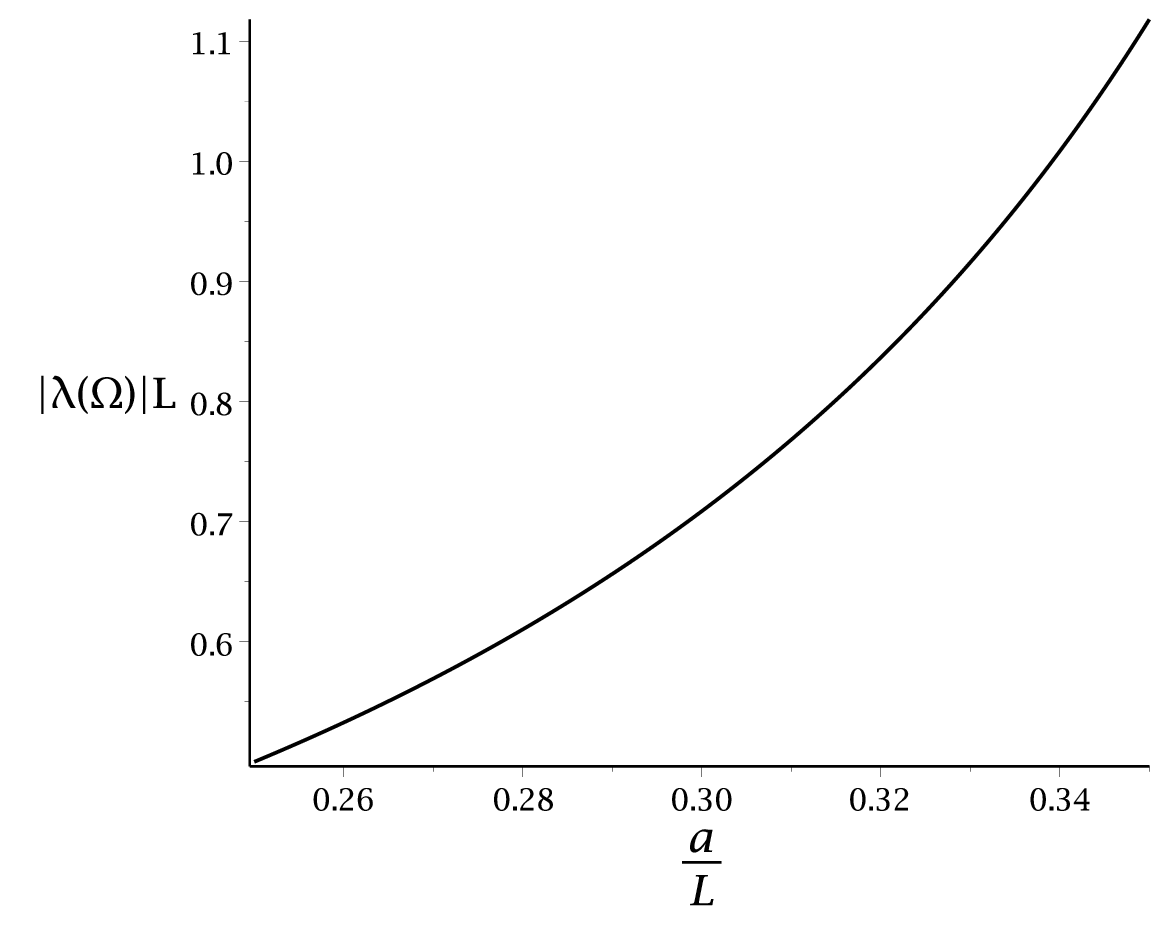}
\caption{ $|\lambda(\Omega)|\big(T^*L,\;\mathfrak{j}/(cL)\big)L$ for AdS$_5$-Kerr, as a function of $a/L$, when $T^*L = 2$ and $r_{<\;\textsf{H}} = 5.5$.}
\end{figure}

Finally, we turn to the most interesting situation, where the temperature is fixed (again, we take $T^*L = 2$ as in Figure 5) but the specific angular momentum is very large. In this case, we see from Figure 1 that we are forced to choose $r_{<\;\textsf{H}}$ to be small in order to be inside the event horizon; let us take $r_{<\;\textsf{H}} = 1$. Then we must take a narrow range of values for $a/L$; we choose 0.94 to 0.98. (Note that $a/L = 0.98$ corresponds to $\mathfrak{j}/(cL) \approx 4.83.$) The result is shown in Figure 8.

\begin{figure}[!h]
\centering
\includegraphics[width=0.70\textwidth]{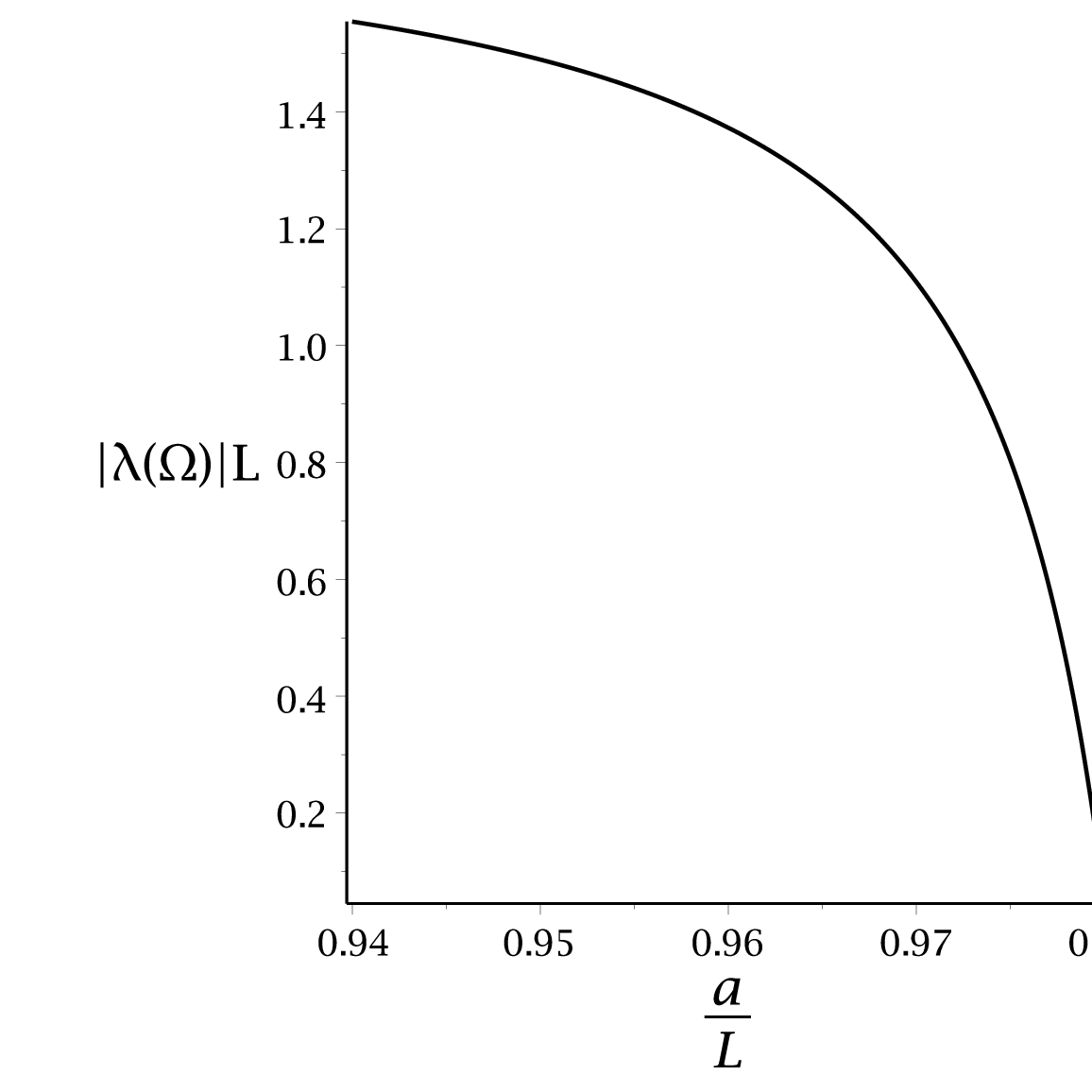}
\caption{ $|\lambda(\Omega)|\big(T^*L,\;\mathfrak{j}/(cL)\big)L$ for AdS$_5$-Kerr, as a function of $a/L$, when $T^*L = 2$ and $r_{<\;\textsf{H}} = 1$.}
\end{figure}

The function rapidly plunges to very small values. Further experimentation with values of $a/L$ even closer to unity shows that in fact $|\lambda(\Omega)|\big(T^*L,\;\mathfrak{j}/(cL)\big)L$ can be driven down to arbitrarily small values by taking $\mathfrak{j}/(cL)$ sufficiently large. Thus we reach the strange conclusion that extremely high specific angular momenta slow down the evolution of the black hole interior just under the horizon, to \emph{any} desired extent. Strange, but not unanticipated: Figure 5 shows exactly this behaviour for $\Gamma\big(T^*L,\;\mathfrak{j}/(cL)\big)$ at large $\mathfrak{j}/(cL)$ (that is, in the same range of values of $a/L$).

In short, then, the numerical evidence rather strongly supports the claim that the rate of change of the interior spatial geometry (just under the horizon) of the AdS$_5$-Kerr black hole does have a holographic dual, in the form of the quantity $\mathfrak{s}T/\hbar$ evaluated on the boundary matter; the duality is expressed by equation (\ref{O}).

\addtocounter{section}{1}
\section* {\large{\textsf{7. Conclusion: Ultra-Vortical Plasmas and Traversability}}}
The observation, at the RHIC facility \cite{kn:STARcoll,kn:STARcoll2}, of high vorticities in the QGP, has attracted a great deal of attention \cite{kn:becca,kn:fei}. The most extreme cases, where the local angular velocity $\omega$ is so large that $c/\omega$ approaches and falls below the mean free path, are now beginning to come into focus, and it is claimed that structures may develop in such plasmas and eventually be observable \cite{kn:buzztuch}. The thermodynamics of ultra-vortical plasmas might well be very unusual and instructive.

We have represented the vortical QGP by a field theory on the conformal boundary of an AdS$_5$-Kerr black hole; the ultra-vortical case is represented dually by such a black hole with a large value of $\mathfrak{j}/(cL)$, where $L$ is the asymptotic curvature scale and $\mathfrak{j}$ is the specific angular momentum of the black hole. We have seen that, if $\mathfrak{j}/(cL)$ is sufficiently large, then our model predicts that the rate of evolution of the boundary theory \emph{slows down}: in Figure 5, this happens once $\mathfrak{j}/(cL)$ exceeds $\approx 1.65,$ which we earlier agreed to call ``large'' in the holographic context. We arrived at a similar result in \cite{kn:110} using a different identification of the physical mass of the black hole (and the technical details are likewise very different), so this conclusion is quite robust.

Our claim is that the dual phenomenon is that, when $\mathfrak{j}/(cL)$ is sufficiently large, wormholes just under the horizon are forced to change slowly (Figure 8). In this model, then, \emph{the existence of slowly evolving wormholes is the characteristic property of the holographic dual of ultra-vortical matter on the boundary.}

An explorer who enters a rapidly rotating AdS$_5$-Kerr black hole will \emph{initially} gain the impression that the spacetime is static, and he might even entertain hopes, on those grounds, of traversing the wormhole. We know, however, that his hopes will be dashed, at least as long as the classical approximation holds: for the null energy condition is satisfied everywhere in this spacetime, so in fact the wormhole is not traversable. This means that, deep inside the black hole (that is, at late times), the geometry becomes dynamic. This in turn means that $\mathfrak{s}\,T/\hbar$ is not a good measure of the rate of evolution of the boundary quantum state at late times. This does not surprise us, since it was clear from the outset that this quantity is too simple to account fully for the boundary dynamics.

A full picture, valid at \emph{all} times, of the duality we are discussing here, would require both a good understanding of the region deep below the event horizon (a notorious problem even at the classical level, for a rotating black hole) and a much more sophisticated measure of the rate of evolution on the boundary. Intuitively, however, the slow \emph{initial} rate at which the internal geometry changes when the specific angular momentum is large suggests that rapidly spinning black holes are ``more nearly traversable'' than their slowly-rotating counterparts.

This is suggestive, since it is now well known that, under certain specific conditions, wormholes may become traversable when quantum backreaction effects are taken into account \cite{kn:aronwall,kn:juan,kn:bilotta}. In particular, in \cite{kn:bilotta} it is argued that this can occur when a rotating black hole is close to extremality. These results have not yet been extended to the AdS$_5$-Kerr case, and the extension is likely to be technically challenging, but there is no reason to think that this cannot be done.

We therefore speculate that \emph{the holographic duals of ultra-vortical plasmas involve quantum-traversable wormholes.} Showing this would involve an extension of the work in \cite{kn:bilotta} to the non-extremal case (since our argument assumes that the Hawking temperature is relatively high).

If this is so, then what does it mean for the ultra-vortical QGP, which might be produced in very high-energy collisions at the LHC \cite{kn:hot}? One of the striking results of \cite{kn:aronwall} was a link between the quantum traversability of a wormhole and (a special form of) quantum teleportation \cite{kn:telep}. It is difficult to see how quantum teleportation is possible in a hot plasma; if it is not, this might lead to useful restrictions.

\addtocounter{section}{1}
\section*{\large{\textsf{Acknowledgement}}}
The author is grateful to Dr. Soon Wanmei for helpful comments.

\end{document}